\documentclass[aps,prl,twocolumn,floatfix,showpacs]{revtex4}
\usepackage[dvips]{color}
\usepackage{amsmath,amssymb,amsfonts,bm,graphicx,latexsym}
\usepackage[hypertex,breaklinks=true,colorlinks=false]{hyperref}

\newcommand{\cutspace}{\vspace{-0.6cm}}

\newcommand{\ua}{\uparrow}
\newcommand{\da}{\downarrow}

\newcommand{\rv}{\bm{r}}

\newcommand{\Kv}{\bm{K}}

\begin{document}

\title{
Integer Quantum Hall State in Two-Component Bose Gases\\ in a Synthetic Magnetic Field
}
\author{Shunsuke Furukawa}
\affiliation{Department of Physics, University of Tokyo, 7-3-1 Hongo, Bunkyo-ku, Tokyo 113-0033, Japan}
\author{Masahito Ueda}
\affiliation{Department of Physics, University of Tokyo, 7-3-1 Hongo, Bunkyo-ku, Tokyo 113-0033, Japan}
\date{\today}


\begin{abstract}
We study two-component (or pseudospin-$1/2$) Bose gases in a strong synthetic magnetic field. 
Using exact diagonalization, 
we show that a bosonic analogue of an integer quantum Hall state with no intrinsic topological order 
appears at the total filling factor $\nu=1+1$ 
when the strengths of intracomponent and intercomponent interactions are comparable with each other. 
This provides a prime example of a symmetry-protected topological phase 
in a controlled setting of quantum gases. 
The real-space entanglement spectrum of this state is found to be comprised of counter-propagating chiral modes 
consistent with the edge theory derived from an effective Chern-Simons theory. 
\end{abstract}
\maketitle



There has been an ever growing interest in artificially created gauge fields in ultracold atomic gases. 
Gauge fields can be induced in neutral atoms by rotating gases \cite{Cooper08_review} or by optically dressing atoms \cite{Lin09,Dalibard11}. 
A sufficiently strong synthetic magnetic field for ultracold atoms is expected to offer interesting analogues of quantum Hall states 
with a rich variety of statistics and internal states of constituent particles. 
The relevant parameter for describing a high magnetic-field regime is the filling factor $\nu=N/N_\phi$, 
where $N$ is the number of atoms and $N_\phi$ is the number of magnetic flux quanta piercing the system. 
For scalar bosons, quantum Hall states have been predicted to appear at various integer and fractional $\nu(\lesssim 6)$ \cite{Cooper08_review}  
such as a bosonic Laughlin state at $\nu=1/2$ \cite{Wilkin98} 
and non-Abelian Read-Rezayi states \cite{Read99} at $\nu=k/2$ with integer $k\ge 2$ \cite{Cooper01}. 


From a general perspective, 
Laughlin and Read-Rezayi states are examples of topologically ordered states
--- in the sense that they host fractional excitations in the bulk and 
exhibit ground-state degeneracies that depend on the system's spatial topology \cite{Wen04_book}. 
Recently, there has been an intense interest in a new type of phases 
which are {\it not} topologically ordered but are distinct from trivial phases as long as certain symmetries are imposed. 
Examples include topological insulators \cite{TopoIns} and Haldane chains \cite{Haldane83_spin}. 
Such phases are termed symmetry-protected topological (SPT) phases. 
Very recently, a systematic classification of SPT phases for interacting bosons has been proposed \cite{Chen12,Lu12}, 
opening up possibilities of hitherto unknown various SPT phases. 
It has been predicted that a system of bosons subject to the particle-number conservation can exhibit 
an integer quantum Hall (IQH) effect with the Hall conductivity quantized to {\it even} integers \cite{Lu12}; 
the resulting states are analogous to the fermionic IQH states in the sense that they both host elementary excitations of unit (rather than fractional) charge.  
It is interesting to ask whether such new SPT phases can be realized in quantum gases with tunable interactions and synthetic gauge fields. 
Quite recently, Senthil and Levin \cite{Senthil13} have proposed that two-component Bose gases in a magnetic field 
may realize one of such phases --- a bosonic IQH state with its Hall conductivity quantized to $2$.  
This intriguing proposal was based on an effective field theory, 
aiming at discussing a general possibility without specifying a microscopic Hamiltonian. 
However, an important issue of the competition among the bosonic IQH state and other candidate states in a realistic Hamiltonian has remained elusive. 


In this Letter, we numerically study a simple realistic system of two-component (or pseudospin-$1/2$) Bose gases in a high magnetic field. 
Recent exact diagonalization studies of this system using a torus geometry \cite{Grass12,FurukawaUeda12} 
have identified a series of incompressible states at $\nu=k/3+k/3$ ($k/3$ filling for each component) with integer $k$
for the case of pseudospin-independent interactions. 
Possible candidate quantum Hall states for these filling factors include non-Abelian spin-singlet states with an $SU(3)_k$ symmetry \cite{Ardonne99}; 
such states have been found to appear for $k=1$ and $2$ \cite{Grass12,FurukawaUeda12}. 
The nature of incompressible ground states (GS) at higher $\nu$ has remained to be identified. 
Here, by combining calculations on torus and sphere geometries, 
we analyze the competition among candidate quantum Hall states. 
We show that the bosonic IQH state of Ref.~\onlinecite{Senthil13} does appear at $\nu=1+1$ 
when the strengths of intracomponent and intercomponent interactions are comparable with each other. 
This result establishes the appearance of an SPT phase in a simple microscopic model, 
and opens up a unique avenue for its realization in a controlled setting of quantum gases. 
We discuss possible experiments for observing this state in light of the current status of creating synthetic gauge fields in ultracold atomic systems.


\begin{figure}
\begin{center}
\includegraphics[width=0.5\textwidth]{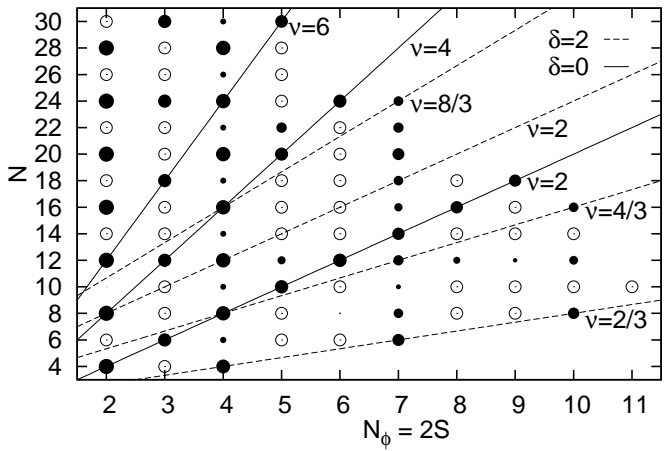}
\end{center}
\cutspace
\caption{
Candidates for incompressible ground states (GSs) in the $(N_\phi,N)$ plane, 
calculated on a sphere geometry in the $SU(2)$-symmetric case $g_{\ua\da}=g$. 
Filled circles indicate GSs with the total angular momentum $L=0$, where incompressible states can appear; 
the area of each symbol is proportional to the neutral gap $\Delta_n$. 
Empty circles indicate the GSs with $L>0$. 
Solid and broken lines show the relation $N = \nu (N_\phi+\delta)$ for $\delta=0$ and $2$, respectively. 
Data points are missing for large $N_\phi$ and $N$ due to an exponentially long computation time. 
Below the dotted line with $\nu=2/3$, GSs are highly degenerate \cite{Paredes02}. 
}
\label{fig:gsL}
\end{figure}


We consider a system of a two-dimensional (2D) Bose gas (in the $xy$ plane) having two hyperfine spin states (labeled by $\alpha=\ua,\da$) 
and subject to a synthetic magnetic field $B$ along the $z$ axis. 
In the case of a rotating system, the magnetic field $B=2M\Omega/q$ is induced in the rotating frame of reference, 
where $M$ is the particle's mass, $\Omega$ is the rotation frequency, and $q$ is a fictitious charge for neutral atoms. 
Unique vortex structures for moderate rotation frequencies 
have been investigated both theoretically \cite{Mueller02,Kasamatsu03} and experimentally \cite{Schweikhard04}. 
Quantum Hall states are expected to emerge as the number of flux quanta (or vortices), $N_\phi$, becomes comparable with the number of atoms, $N$, 
for sufficiently strong $B$. 
For such large $B$, it is natural to assume that the atomic motion in the $xy$ plane is restricted to the lowest Landau level (LLL). 
Within this subspace, 
we consider the interaction Hamiltonian consisting of intracomponent and intercomponent contact interactions: 
\begin{equation}\label{eq:Hint}
 H_\mathrm{int} 
 = g \sum_{\alpha=\ua,\da} \sum_{i<j}^{N_\alpha} \delta (\rv_i^\alpha-\rv_j^\alpha) 
 + g_{\ua\da}  \sum_{i=1}^{N_\ua} \sum_{j=1}^{N_\da} \delta (\rv_i^\ua-\rv_j^\da) , 
\end{equation}
where $N_\alpha$ is the number of particles in state $\alpha$,  
$\rv_i^\alpha$'s are the 2D positions of the particles, and 
$g$ and $g_{\ua\da}$ are the effective intracomponent and intercomponent coupling constants in the 2D plane. 
We mainly focus on the case of $SU(2)$-symmetric interactions $g_{\ua\da}=g$; 
the case of $g_{\ua\da}\ne g$ will also be discussed. 




To study bulk properties, 
it is useful to work on closed uniform manifolds having no edge. 
In Refs.~\onlinecite{Grass12} and \onlinecite{FurukawaUeda12}, 
a periodic rectangular geometry (torus) of sides $L_x$ and $L_y$ is used. 
Here, we also use a sphere geometry and compare the two geometries to identify geometry-independent robust features. 
For a sphere geometry \cite{Haldane83}, 
a magnetic monopole of charge $N_\phi (2\pi \hbar/q)$ with integer $N_\phi\equiv 2S$ is placed at the center, 
and it produces a uniform magnetic field $B$ on the sphere of radius $R=\ell \sqrt{S}$, 
where $\ell=\sqrt{\hbar/(q B)}$ is the magnetic length. 
Introducing the spherical coordinates $(\theta,\varphi)$, 
single-particle orbits in the LLL are given by 
$\psi_m\propto (v^*)^{S+m} (-u^*)^{S-m}$ with $u=\cos(\theta/2)e^{i\varphi/2}$ and $v=\sin(\theta/2)e^{-i\varphi/2}$, 
where $m=-S,-S+1,\dots,S$ is the $z$-component of the angular momentum. 
On both the sphere and the torus, the filling factor in the thermodynamic limit 
is given by $\nu=N/N_\phi$ with $N=N_\ua+N_\da$. 
On a finite sphere, however, 
the relation between $N$ and $N_\phi$ involves a characteristic shift $\delta$ for incompressible states:
$N = \nu (N_\phi+\delta)$, 
where $\delta$ depends on an individual candidate wave function. 
Therefore, on a sphere, competing incompressible states leading to the same $\nu$ in the thermodynamic limit 
can be studied separately with different $(N_\phi, N)$ if they have different shifts. 
On a torus, there is no shift, and all candidates for the same $\nu$ compete in the same finite-size calculation. 
While a torus can give less biased results, 
a sphere is more useful in discussing relative stabilities of given candidates. 
Henceforth, we take the units of $\hbar\equiv 1$ and $\ell\equiv 1$. 


In Fig.~\ref{fig:gsL}, we set $g_{\ua\da}=g>0$ and look for candidates of quantum Hall states in the $(N_\phi, N)$ plane using a sphere geometry. 
Because of the rotational symmetry of a sphere, many-body eigenstates  
can be classified by the total angular momentum $L$. 
It is known that incompressible states in general appear as unique GSs with $L=0$, 
which are indicated by filled circles in Fig.~\ref{fig:gsL}. 
The area of each filled circle is proportional to the neutral gap $\Delta_n$ to be defined later. 
Solid and broken lines indicate the relation $N = \nu (N_\phi+\delta)$ for $\delta=0$ and $2$, respectively. 
The bosonic IQH state with $\nu=2$ has $\delta=0$ \cite{Senthil13}; 
the $SU(3)_k$ states with $\nu=2k/3$ have $\delta=2$ and can appear only if $N_\phi+2$ is a multiple of $3$ \cite{Ardonne99}. 
We find that the $L=0$ GSs appear for all $(N_\phi,N)$ corresponding to the IQH and $SU(3)_k$ states. 
Furthermore, we also find $L=0$ GSs with appreciable energy gaps along the $\delta=0$ lines with $\nu=4$ and $6$, 
whose nature will be discussed later. 
Below we analyze the stabilities of these candidate states and, in particular, the competition between the IQH state and $SU(3)_3$ state at $\nu=2$ 
based on the data of energy gaps. 


\begin{figure}
\begin{center}
\includegraphics[width=0.5\textwidth]{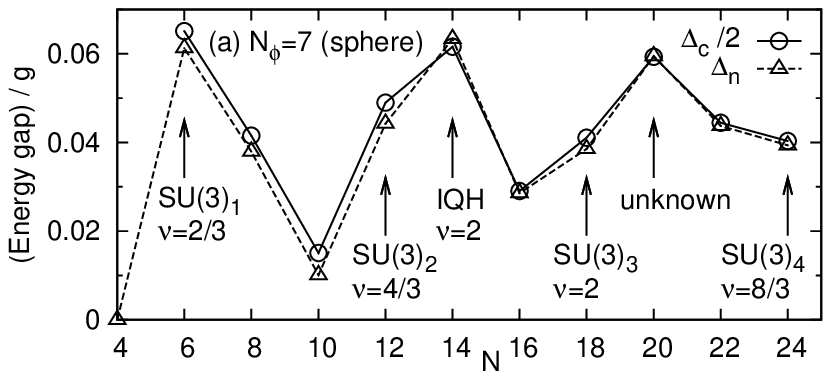}\\
\includegraphics[width=0.5\textwidth]{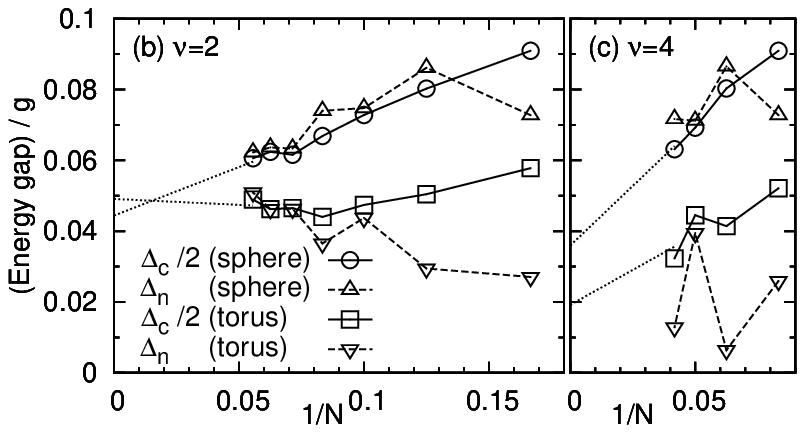}
\end{center}
\cutspace
\caption{
(a) Charge and neutral energy gaps, $\Delta_c$ and $\Delta_n$, calculated on a sphere with $N_\phi=7$. 
(b,c) Size dependences of the energy gaps at $\nu=2$ and $4$ on the sphere and torus geometries. 
For a torus, we set $L_x=L_y$ (periodic square). 
For a sphere, we set $\delta=0$; 
namely, calculations are performed along the solid lines in Fig.~\ref{fig:gsL}. 
The dotted lines indicate linear extrapolation of the charge gap to the thermodynamic limit, using the data with $N\ge 10$. 
}
\label{fig:Gap}
\end{figure}


We consider two types of energy gaps, namely, charge and neutral gaps. 
The charge gap is defined as 
$\Delta_c (N) = E_0\left( N/2+1, N/2 \right) + E_0\left( N/2-1, N/2 \right) - 2 E_0\left( N/2, N/2 \right) $
for even integer $N(\ge4)$, 
where $E_0(N_\ua,N_\da)$ is the ground-state energy for given $(N_\ua,N_\da)$. 
The neutral gap $\Delta_n$ is defined as the gap to the first excited state for $(N_\ua,N_\da)=(N/2,N/2)$. 


Figure~\ref{fig:Gap}(a) presents the result for a sphere with $N_\phi=7$ and varying $N$. 
Both the IQH state and the $SU(3)_k$ states can appear in this case. 
The two types of gaps, $\Delta_c/2$ and $\Delta_n$, show essentially similar behaviors 
[the former is divided by $2$ because of a doubled energy scale involved in the definition of $\Delta_c$]. 
The most prominent gap is found at $N=6$, 
where the Halperin $(221)$ state \cite{Halperin83} [or the $SU(3)_1$ state] gives the exact GS \cite{Paredes02}. 
The second largest gap (comparable with the first) is found at $N=14$, the latter corresponding to the $\nu=2$ IQH state. 
This gap is (by a factor of about $1.5$) larger than the gap at $N=18$, which corresponds to the $SU(3)_3$ state. 
This indicates that, of the two candidates at $\nu=2$, the IQH state is likely to be more stable. 
By comparing Fig.~\ref{fig:Gap}(a) with the previous result on a torus \cite{Grass12,FurukawaUeda12}, 
where comparable gaps at $\nu=2/3$ and $2$ were found, 
we find that interpreting the $\nu=2$ GS as the IQH state gives a better consistency between the sphere and torus data. 
In Fig.~\ref{fig:Gap}(a), we also find a large gap at $N=20$ (comparable with the two largest), whose physical origin is unclear. 


In Fig.~\ref{fig:Gap}(b) and (c), we examine size dependences of the gaps at $\nu=2$ and $4$ on the sphere and torus geometries. 
A sphere tends to give larger gaps than a torus for finite sizes. 
For $\nu=2$ in Fig.~\ref{fig:Gap}(b), we find that the difference between the sphere and torus data decreases as $N$ increases. 
Extrapolation to the thermodynamic limit (dotted lines) yields 
$\Delta_c/(2 g)\approx 0.44$ and $0.49$ for the sphere and torus cases; 
they are in good agreement, given the residual oscillations in finite-size data used for the extrapolation. 
Furthermore, the torus data give an additional support for excluding the appearance of the $SU(3)_3$ state. 
If the $SU(3)_3$ state is stabilized in the thermodynamic limit, 
$\Delta_n$ in the present definition should drop 
because of the $10$-fold topological GS degeneracy expected on a torus \cite{Ardonne99} 
(a real neutral gap should be found above the degenerate GSs in this case). 
We find no appreciable drops of $\Delta_n$ for $N=6,12,18$ (system sizes where the $SU(3)_3$ state is allowed) in Fig.~\ref{fig:Gap}(b). 
For $\nu=4$ in Fig.~\ref{fig:Gap}(c), the system size is not large enough to check the consistency between the torus and sphere data; 
yet, simple linear extrapolation of the charge gap (dotted lines) indicates a tendency toward a nonzero value in the thermodynamic limit on both geometries. 



For $\nu=2$, we have also analyzed the case of $g_{\ua\da}\ne g$ by calculating the spectra on a torus \cite{Suppl}. 
We find that the IQH state with an appreciable energy gap survives against moderate changes in the ratio $g_{\ua\da}/g$. 

\begin{figure}
\begin{center}
\includegraphics[width=0.5\textwidth]{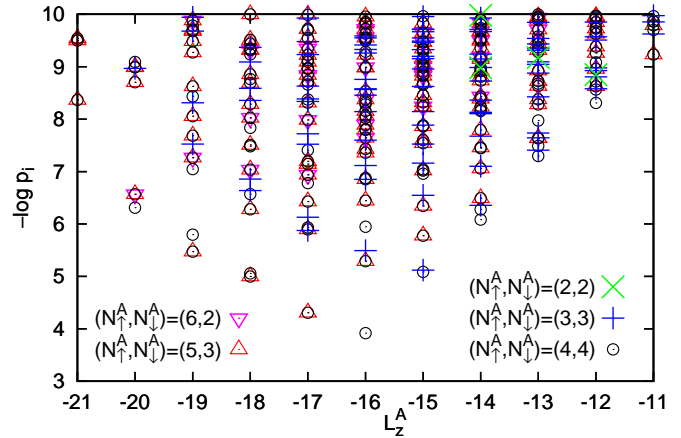}
\end{center}
\cutspace
\caption{(Color online) 
Real-space entanglement spectrum of the $\nu=2$ IQH state with $(N_\phi,N)=(8,16)$ on a sphere. 
We plot the entanglement energies $\{-\log p_i\}$ 
defined from the eigenvalues $\{p_i\}$ of $\rho_A$, where $A$ is the northern hemisphere; 
the lower the entanglement energy is, the larger weight it has in $\rho_A$. 
The entanglement energies are classified by the number of $\ua$ and $\da$ atoms, $N_\ua^A$ and $N_\da^A$, 
and by the $z$-component of the total angular momentum, $L_z^A$, on $A$. 
}
\label{fig:ES}
\end{figure}

We have so far discussed the appearance of the IQH state at $\nu=2$ using only spectral properties. 
A useful way to unveil the topological features of this state is to investigate the entanglement spectrum \cite{Li08}. 
This is an eigenvalue spectrum of a reduced density matrix $\rho_A$ on a subsystem $A$, 
and has been demonstrated to serve as a useful probe of the edge spectrum. 
Here we use the GS on a sphere, split it into two hemispheres, 
and calculate the real-space entanglement spectrum (RSES) \cite{Sterdyniak12} for the northern hemisphere $A$. 
The result is presented in Fig.~\ref{fig:ES}. 
Below we analyze this result based on an effective field theory. 

The $\nu=2$ bosonic IQH state is described by the Chern-Simons theory with the $2\times 2$ $K$ matrix with 
$K_{12}=K_{21}=1$ and $K_{11}=K_{22}=0$ \cite{Lu12,Senthil13}. 
Through the standard procedure \cite{Wen04_book}, 
the effective Lagrangian for one-dimensional edge modes is obtained as
\begin{equation}
 {\cal L}= -\frac{1}{4\pi} (K_{\alpha\beta} \partial_t\phi_\alpha \partial_x\phi_\beta + V_{\alpha\beta} \partial_x\phi_\alpha \partial_x\phi_\beta), 
\end{equation}
where $\frac{1}{2\pi} \partial_x\phi_\alpha$ gives the density of bosons of spin $\alpha=\ua,\da$ 
and $V_{\alpha\beta}$ is the velocity matrix. 
Diagonalizing ${\cal L}$ via charge and spin modes 
$\phi_{c/s}=(\phi_\ua \pm \phi_\da)/\sqrt{2}$, 
one finds that the charge (spin) mode is right-moving (left-moving) \cite{Senthil13}.  
Introducing the velocities $v_{c/s}>0$ for these modes and performing mode expansions, 
the Hamiltonian and the total momentum for the edge of length $L_x$ are diagonalized as \cite{Suppl}
\begin{align}\label{eq:H_edge}
 H= \frac{2\pi}{L_x} (v_c L_0^c + v_s L_0^s), 
~P= \frac{2\pi}{L_x} (L_0^c - L_0^s),  
\end{align}
with
\begin{align}\label{eq:edge_spec}
L_0^{c/s} = \frac{(\Delta N_\ua\pm \Delta N_\da)^2}{4} + \sum_{m=1}^\infty \! m n_{m}^{c/s} . 
\end{align}
Here, $\Delta N_\alpha$ ($\alpha=\ua,\da$) is a change in the number of spin-$\alpha$ bosons 
relative to the GS; 
$\{n_m^c\}$ and $\{n_m^s\}$ are sets of non-negative integers describing oscillator modes. 
The second term in Eq.~\eqref{eq:edge_spec} exhibits the well-known counting of degeneracies 
associated with a $U(1)$ free boson: $1,1,2,3,\dots$. 

In Fig.~\ref{fig:ES}, the lowest entanglement energy is found at $L_z^A=-16$ with $(N_A^\ua,N_A^\da)=(4,4)$, 
which can be identified with the ``GS'' of virtual edge modes probed by the RSES. 
``Excited states'' in the RSES are found in both positive and negative directions of $L_z^A$ relative to the ``GS'', 
which indicates the counter-propagating nature of the edge modes. 
This is in sharp contrast to the case of a single chiral edge mode \cite{Li08,Sterdyniak12}, 
where low-lying ``excited states'' are found only in one direction of $L_z^A$. 
The excitations along the left envelope of the spectrum can be interpreted as the left-moving spin mode with $P<0$, 
as spin excitations with $(\Delta N_A^\ua, \Delta N_A^\da)=(+1,-1)$ and $(+2,-2)$ relative to the ``GS'' are found there; 
the shifts $\Delta L_z^A=-1$ and $-4$ of these excitations are consistent with Eqs.~\eqref{eq:H_edge} and \eqref{eq:edge_spec}. 
Similarly, the excitations along the right envelope can be interpreted as the right-moving charge mode with $P>0$, 
as charge excitations with $(\Delta N_A^\ua, \Delta N_A^\da)=(-1,-1)$ and $(-2,-2)$ are found there. 
The levels appearing between the two modes can be identified with combinations of these modes. 


Let us briefly discuss the nature of the gapped states at $\nu=4,6,\dots$ found in Fig.~\ref{fig:gsL}. 
The RSES of the $\nu=4$ state reveals a counter-propagating nature of edge modes \cite{Suppl}, as in the $\nu=2$ case in Fig.~\ref{fig:ES}. 
This suggests that the $\nu=4$ state has certain similarities with the $\nu=2$ IQH state. 
However, the bosonic IQH states with $\nu\ge 4$ predicted in Ref.~\cite{Lu12} are {\it not} good candidates for the present system, 
as their corresponding simplest $K$ matrices are not symmetric with respect to the interchange of the two components. 
A more detailed characterization of the states at $\nu=4,6,\dots$ is left for future studies. 


Finally, we discuss possible experiments for observing the IQH state found in this Letter. 
In ultracold atom experiments, a clear signature of a quantum Hall state would be a density plateau in an {\it in situ} image of a gas. 
A challenge here is how to realize a large synthetic field required for observing this state. 
In Ref.~\cite{Lin09}, by using an optical coupling between $F=1$ internal states of $^{87}$Rb atoms with space-dependent detuning, 
a flux density $(2\pi\ell^2)^{-1} \approx 0.076 ~\mu$m$^{-2}$ has been realized. 
This produces the Landau-level spacing $\Delta_\mathrm{LL}\approx 2.7$ nK$\times k_B$. 
It is an important technical challenge to increase this field by a factor of a few tens 
so that $\Delta_\mathrm{LL}$ dominates the experimental temperature scales. 
This could be achieved by further increasing the detuning gradient in the scheme of Ref.~\onlinecite{Lin09} 
or by using spatially shifted laser beams \cite{Dalibard11}. 
We hope that the appearance of exotic phases demonstrated in this Letter and Refs.~\onlinecite{Grass12} and \onlinecite{FurukawaUeda12} 
stimulate experimental attempts to create such large synthetic magnetic field in two-component Bose gases, 
by, for example, separately creating synthetic fields in the $F=1$ and $2$ states of $^{87}$Rb. 
Once such a large field is created, a quantum Hall regime with $\nu=O(1)$ is expected to be realized 
by splitting the gas into an array of 2D systems by using a one-dimensional optical lattice along the $z$ direction as suggested in Ref.~\cite{Lin09}. 
By strongly squeezing each 2D system along the $z$ direction, the IQH state exhibits a reasonable gap for its observation \cite{Suppl}. 








{\it Note added.} 
After the submission of this Letter, 
we became aware of two independent works \cite{Wu13,Regnault13}, 
which also study the ground state at $\nu=2$ and have some overlap with this Letter \cite{Suppl}. 


This work was supported 
by KAKENHI Grant No. 25800225 from Japan Society for the Promotion of Science, 
and by a Grant-in-Aid for Scientific Research on Innovative Areas ``Topological Quantum Phenomena'' (KAKENHI No. 22103005), 
KAKENHI Grant No. 22340114, 
a Global COE Program ``the Physical Science Frontier'', 
and the Photon Frontier Network Program 
from MEXT of Japan. 

\newcommand{\etal}{{\it et al.}}
\newcommand{\PRL}[3]{Phys. Rev. Lett. {\bf #1}, \href{http://link.aps.org/abstract/PRL/v#1/e#2}{#2} (#3)}
\newcommand{\PRLp}[3]{Phys. Rev. Lett. {\bf #1}, \href{http://link.aps.org/abstract/PRL/v#1/p#2}{#2} (#3)}
\newcommand{\PRA}[3]{Phys. Rev. A {\bf #1}, \href{http://link.aps.org/abstract/PRA/v#1/e#2}{#2} (#3)}
\newcommand{\PRAp}[3]{Phys. Rev. A {\bf #1}, \href{http://link.aps.org/abstract/PRA/v#1/p#2}{#2} (#3)}
\newcommand{\PRAR}[3]{Phys. Rev. A {\bf #1}, \href{http://link.aps.org/abstract/PRA/v#1/e#2}{#2} (R) (#3)}
\newcommand{\PRB}[3]{Phys. Rev. B {\bf #1}, \href{http://link.aps.org/abstract/PRB/v#1/e#2}{#2} (#3)}
\newcommand{\PRBp}[3]{Phys. Rev. B {\bf #1}, \href{http://link.aps.org/abstract/PRB/v#1/p#2}{#2} (#3)}
\newcommand{\PRBR}[3]{Phys. Rev. B {\bf #1}, \href{http://link.aps.org/abstract/PRB/v#1/e#2}{#2} (R) (#3)}
\newcommand{\RMP}[3]{Rev. Mod. Phys. {\bf #1}, \href{http://link.aps.org/abstract/RMP/v#1/e#2}{#2} (#3)}
\newcommand{\arXiv}[1]{arXiv:\href{http://arxiv.org/abs/#1}{#1}}
\newcommand{\condmat}[1]{cond-mat/\href{http://arxiv.org/abs/cond-mat/#1}{#1}}
\newcommand{\JPSJ}[3]{J. Phys. Soc. Jpn. {\bf #1}, \href{http://jpsj.ipap.jp/link?JPSJ/#1/#2/}{#2} (#3)}
\newcommand{\PTPS}[3]{Prog. Theor. Phys. Suppl. {\bf #1}, \href{http://ptp.ipap.jp/link?PTPS/#1/#2/}{#2} (#3)}
\newcommand{\hreflink}[1]{\href{#1}{#1}}


\appendix

\section{Supplemental Material}

\subsection{Stability of the integer quantum Hall state at $\nu=2$}

\begin{figure}[b]
\begin{center}
\includegraphics[width=0.50\textwidth]{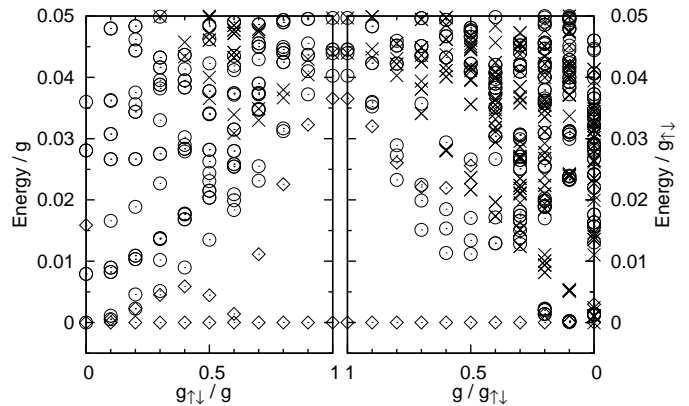}
\cutspace
\end{center}
\caption{
Energy spectra versus coupling ratio $g_{\ua\da}/g$ (left) and its inverse (right) for $(N_\phi,N)=(6,12)$. 
Diamonds ($\Diamond$) indicate two lowest-energy states at $\Kv=0$ in the equal-population case $N_\ua=N_\da=N/2$; 
the ground-state energy of this sector is subtracted from the whole spectrum.  
Circles ($\bigcirc$) indicate other eigenstates in the equal-population case. 
Crosses ($\times$) show eigenstates for the minimally imbalanced case of $N_\ua=N/2+1$ and $N_\da=N/2-1$. 
}
\label{fig:ener_g}
\end{figure}

Here we consider the case of $g_{\ua\da}\ne g$ 
and discuss the stability of the integer quantum Hall (IQH) state found at $\nu=2$. 
Figure~\ref{fig:ener_g} presents the spectra for $(N_\phi,N)=(6,12)$ on a torus geometry 
as a function of the coupling ratio $g_{\ua\da}/g$ or its inverse. 
We find that an appreciable gap above a non-degenerate ground state remains 
in the regions $0.7 \lesssim g_{\ua\da}/g \le 1$ (left panel) and $1 \ge g/g_{\ua\da} \gtrsim 0.3$ (right panel). 
This indicates that the IQH state is stable against moderate changes in the ratio $g_{\ua\da}/g$ around the $SU(2)$-symmetric case. 

\subsection{Pair distribution functions}

\begin{figure}
\begin{center}
\includegraphics[width=0.50\textwidth]{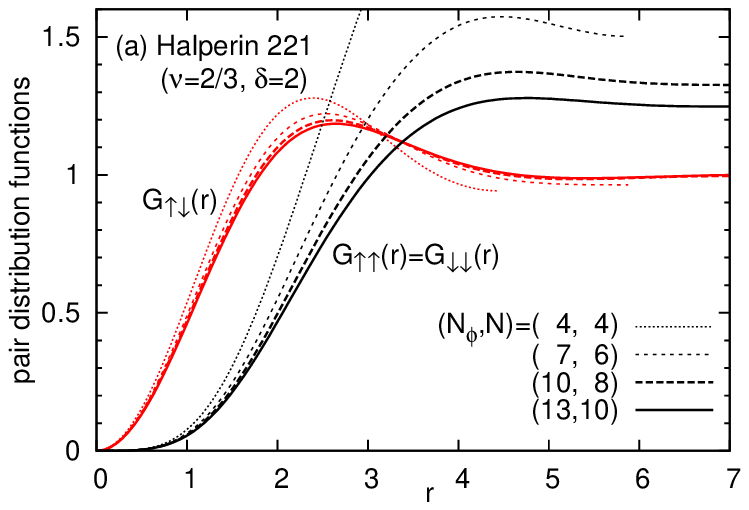}
\includegraphics[width=0.50\textwidth]{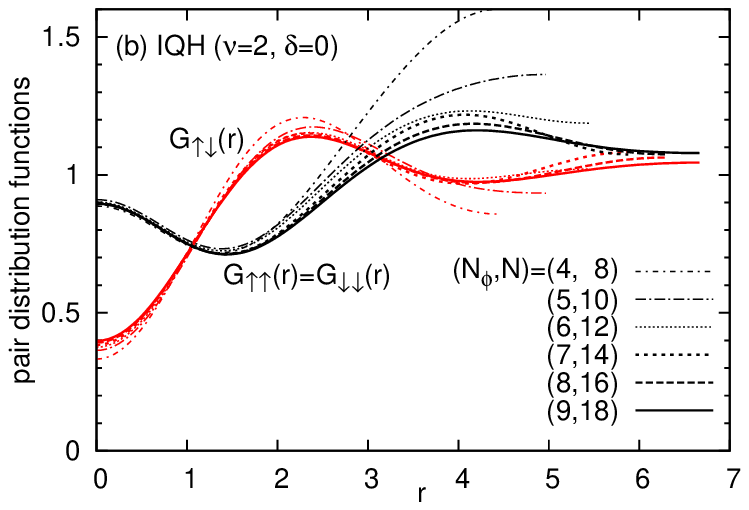}
\cutspace
\end{center}
\caption{
Pair distribution functions \eqref{eq:pdf} plotted against $r=|\rv|$ 
for (a) the Halperin $(221)$ state at $\nu=2/3$ and (b) the bosonic IQH state at $\nu=2$. 
}
\label{fig:pdf_all}
\end{figure}

Pair distribution functions provide a useful probe of particle correlations 
present in the ground-state wave functions. 
For a uniform system of area $A$, they are defined as
\begin{align} \label{eq:pdf}
&G_{\alpha\alpha} (\rv)= \frac{A}{N_\alpha (N_\alpha-1)} \bigg\langle \sum_{i\ne j} \delta(\rv+\rv_i^\alpha-\rv_j^\alpha) \bigg\rangle,~
\alpha=\ua,\da,\\
&G_{\ua\da} (\rv)= \frac{A}{N_\ua N_\da} \bigg\langle \sum_{i, j} \delta(\rv+\rv_i^\ua-\rv_j^\da) \bigg\rangle.
\end{align}
We have calculated these functions using the ground states obtained by exact diagonalization on a sphere geometry. 
The results for the Halperin $(221)$ state at $\nu=4/3$
and the bosonic IQH state at $\nu=2$ are presented in Fig.~\ref{fig:pdf_all}. 
On a disc geometry, the Halperin $(221)$ state is given by \cite{Halperin83S}
\begin{equation}
 \Psi^{221} = \prod_{i<j} (z_i^\ua-z_j^\ua)^2 \prod_{i<j} (z_i^\da-z_j^\da)^2 \prod_{i,j} (z_i^\ua-z_j^\da) 
 e^{-\sum_{i,\alpha} |z_i^\alpha|^2/4}, 
\end{equation}
where $z_i^\alpha=x_i^\alpha+iy_i^\alpha$ is a complex coordinate. 
In this wave function, the pair distribution functions obey peculiar power laws 
$G_{\alpha\alpha}(r)\propto r^4$ and $G_{\ua\da}(r)\propto r^2$ as $r \to 0$, 
which are also found in the data in Fig.~\ref{fig:pdf_all}(a). 
By contrast, the wave function for the $\nu=2$ IQH state has been predicted to be \cite{Senthil13S} 
\begin{equation}\label{eq:Psi_IQH}
\begin{split}
 \Psi^\mathrm{IQH} = P_\mathrm{LLL} &\prod_{i<j} |z_i^\ua-z_j^\ua|^2 \prod_{i<j} |z_i^\da-z_j^\da|^2 \\
 &\times \prod_{i,j} (z_i^\ua-z_j^\da) 
 e^{-\sum_{i,\alpha} |z_i^\alpha|^2/4}, 
\end{split}
\end{equation}
where $P_\mathrm{LLL}$ is the projection onto the lowest Landau-level manifold. 
While the unprojected part of this wave function has the same correlation properties as $\Psi^{221}$, 
the result presented in Fig.~\ref{fig:pdf_all}(b) is very different from Fig.~\ref{fig:pdf_all}(a).  
In particular, we find a hump in $G_{\alpha\alpha}(r)$ for small $r$. 
These results suggest that the correlation properties of the $\nu=2$ state are very different from 
what is expected from the unprojected part of Eq.~\eqref{eq:Psi_IQH}. 

\subsection{Entanglement spectra}

\begin{figure}
\begin{center}
\includegraphics[width=0.50\textwidth]{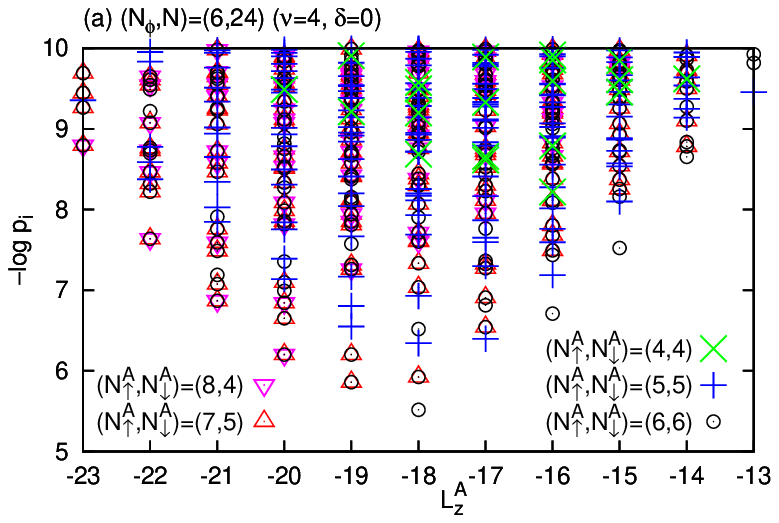}
\includegraphics[width=0.50\textwidth]{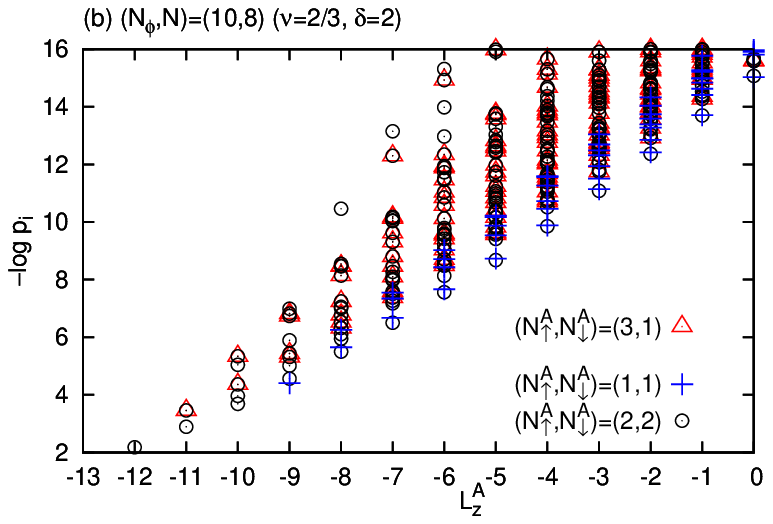}
\cutspace
\end{center}
\caption{
Real-space entanglement spectra for $(N_\phi,N)=(6,24)$ and $(10,8)$ on the sphere geometry. 
We plot the entanglement energies $\{-\log p_i\}$ 
defined from the eigenvalues $\{p_i\}$ of $\rho_A$, where $A$ is the northern hemisphere.  
The entanglement energies are classified by the numbers of $\ua$ and $\da$ atoms, $N_\ua^A$ and $N_\da^A$, 
and by the $z$-component of the total angular momentum, $L_z^A$, on the northern hemisphere. 
}
\label{fig:ES_all}
\end{figure}

Figure~\ref{fig:ES_all} presents the real-space entanglement spectra (RSES) 
of candidate incompressible states on a sphere geometry.  
Here we divide the sphere into two hemispheres, 
calculate the reduced density matrix $\rho_A$ on the northern hemisphere $A$, 
and plot the entanglement energies $\{-\log p_i\}$ defined from the eigenvalues $\{p_i\}$ of $\rho_A$. 
The case of $(N_\phi,N)=(6,24)$ in Fig.~\ref{fig:ES_all}(a) 
corresponds to the incompressible state with $\nu=4$ and $\delta=0$, 
whose nature has been elusive. 
Similarly to the $\nu=2$ case in Fig.~3 of the main text, 
the low-lying entanglement levels appear in both positive and negative directions of $L_z^A$, 
indicating the counter-propagating nature of edge modes. 
This suggests that the $\nu=4$ state has certain similarities with the $\nu=2$ IQH state. 
For comparison, Fig.~\ref{fig:ES_all}(b) presents the RSES of the Halperin $(221)$ state [or $SU(3)_1$ state] at $\nu=2/3$. 
The low-lying levels appear only in the positive direction of $L_z^A$, 
indicating the chiral (right-moving) nature of edge modes of this state. 

\subsection{Derivation of the edge-mode spectrum}

Here we derive the edge-mode spectrum of the $\nu=2$ bosonic IQH state, 
starting from the effective Lagrangian density \cite{Senthil13S}
\begin{equation}
 {\cal L}= -\frac{1}{4\pi} (K_{\alpha\beta} \partial_t\phi_\alpha \partial_x\phi_\beta + V_{\alpha\beta} \partial_x\phi_\alpha \partial_x\phi_\beta), 
\end{equation}
with 
$
K=
\begin{pmatrix}
 0 & 1\\
 1 & 0
\end{pmatrix}
$. 
Here, $\frac{1}{2\pi} \partial_x\phi_\alpha$ gives the density of bosons of spin $\alpha=\ua,\da$   
and $V_{\alpha\beta}$ is the velocity matrix. 
Introducing the charge and spin modes $\phi_{c/s}=(\phi_\ua \pm \phi_\da)/\sqrt{2}$, 
the above Lagrangian density is diagonalized as
\begin{equation}\label{eq:Lcs}
\begin{split}
 {\cal L} = 
 &-\frac{1}{4\pi} (\partial_t\phi_c \partial_x\phi_c + v_c \partial_x\phi_c \partial_x\phi_c)\\
 &-\frac{1}{4\pi} (-\partial_t\phi_s \partial_x\phi_s + v_s \partial_x\phi_s \partial_x\phi_s),  
\end{split}
\end{equation}
where $v_{c/s}(>0)$ are the velocities of these modes, which are obtained as the eigenvalues of the velocity matrix. 
It follows from the Euler-Lagrange equations that the charge mode is right-moving, while the spin mode is left-moving \cite{Senthil13S}:
\begin{equation}
 \phi_c=\phi_c (x-v_c t), ~~
 \phi_s=\phi_s (x+v_s t). 
\end{equation}
From Eq.~\eqref{eq:Lcs}, the Hamiltonian and the total momentum are calculated as
\begin{subequations}\label{eq:HP_cs}
\begin{align}
 H&=\int_0^{L_x} \frac{dx}{4\pi} \left[v_c (\partial_x\phi_c)^2 + v_s (\partial_x\phi_s)^2 \right], \\
 P&=\int_0^{L_x} \frac{dx}{4\pi} \left[ (\partial_x\phi_c)^2 - (\partial_x\phi_s)^2 \right]. 
\end{align}
\end{subequations}
We perform mode expansions for the bosonic fields:
\begin{align}
 \phi_c(x) =& \frac{\Delta N_\ua+ \Delta N_\da}{\sqrt{2}} \frac{2\pi x}{L_x} 
  +\sum_{m=1}^\infty \frac{1}{\sqrt{m}} \left( a_m^c e^{ik_m x} + \mathrm{h.c.} \right),\\
 \phi_s(x) =& \frac{\Delta N_\ua-\Delta N_\da}{\sqrt{2}} \frac{2\pi x}{L_x} 
  +\sum_{m=1}^\infty \frac{1}{\sqrt{m}} \left( a_m^s e^{-ik_m x} + \mathrm{h.c.} \right),
\end{align}
where $\Delta N_\alpha$ ($\alpha=\ua,\da$) is a change in the number of spin-$\alpha$ bosons relative to the GS 
and $a_{m}^{c/s}$ are bosonic operators describing oscillator modes.  
Using these, each term in Eq.~\eqref{eq:HP_cs} can be diagonalized as
\begin{align}
 \int_0^{L_x} \frac{dx}{4\pi} (\partial_x\phi_{c/s})^2
 = \frac{2\pi}{L_x} \left[ \frac{(\Delta N_\ua\pm \Delta N_\da)^2}{4} + \sum_{m=1}^\infty \! m n_{m}^{c/s} \right],  
\end{align}
with $n_m^{c/s}=a_m^{c/s\dagger} a_m^{c/s}$, 
where we have ignored unimportant constant terms.  

\subsection{Evaluation of the excitation gap}

Suppose that we have a large synthetic magnetic field with $(2\pi\ell^2)^{-1} \approx 4.0~\mu$m$^{-2}$ 
for a two-component Bose gas. 
This gives the Landau-level spacing of $\Delta_\mathrm{LL}\approx 140$ nK$ \times k_B$. 
Using a one-dimensional optical lattice along the $z$ direction, 
we split the gas into an array of 2D systems. 
For $^{87}$Rb, the coupling constants for the contact interactions are almost pseudospin-independent: 
\begin{equation}
g^\mathrm{(3D)}\approx g_{\ua\da}^\mathrm{(3D)} \approx \frac{4\pi \hbar^2 a}{M},  
\end{equation}
with the atomic mass $M\approx 1.44\times 10^{-25}$ kg and the $s$-wave scattering length $a\approx 5.5$ nm. 
The effective coupling constants in the 2D plane are given by 
\begin{equation}
 g = \frac{ g^\mathrm{(3D) }}{ \sqrt{2\pi}d_z }  \approx \frac{0.15 \mathrm{nK}\cdot\mu\mathrm{m}^3 \times k_B}{d_z},
\end{equation}
where $d_z$ is the width of each 2D system along the $z$ direction.  
With $d_z=20$ nm and the above flux density, 
the energy gap of the $\nu=2$ IQH state in Fig.~2(b) of the main text gives $\Delta_n\approx 0.05 g/\ell^2\approx 10$ nK$\times k_B$, 
which is a reasonable scale for observing the predicted state in ultracold atom experiments. 

\subsection{Comments on related works}

Here we comment on two related works \cite{Wu13S,Regnault13S}, 
which also study the ground state at $\nu=2$. 
These works and the present one complement one another to some extent, as we describe below. 
Wu and Jain \cite{Wu13S} have calculated the RSES on the sphere and the edge spectrum on the disk, 
and found the presence of counter-propagating modes. 
Figure~3 in the main text agrees with this work. 
They have also constructed a trial wave function, which has a large overlap with the $\nu=2$ ground state. 
Regnault and Senthil \cite{Regnault13S} have analyzed the scaling of the energy gap on the torus 
and furthermore discussed the phase diagram as a function of $g_{\ua\da}/g$. 
Figure~2(b) in the main text agrees with this work. 
Our work employs both the sphere and torus geometries and 
present solid evidences for the IQH state through the consistency of the two geometries. 

Reference \onlinecite{Wu13S} has also constructed a composite fermion wave function for $\nu=4/3$, 
and found a large overlap with the ground state on a sphere with the appropriate shift $\delta=-1$.  
For the same filling factor, Refs.~\onlinecite{Grass12S} and \onlinecite{FurukawaUeda12S} have discussed 
the appearance of the $SU(3)_2$ state on the basis of spectra on a torus geometry. 
Since the composite fermion state and the $SU(3)_2$ state have different shifts, 
they appear at different $(N_\phi,N)$ on the sphere geometry. 
As seen in Fig.~2(a) of the main text, the $SU(3)_2$ state at $N=12$ 
has slightly larger gaps than the composite fermion state at $N=8$. 
A similar result is also obtained for $N_\phi=10$ (not shown). 
This suggests that the $SU(3)_2$ state is likely to be more stable; 
however, the gap values of the two states are very close, 
and calculations for larger system sizes are required to obtain a conclusive result.

\end{document}